\documentclass[pdflatex,sn-mathphys-num]{sn-jnl}

\usepackage{graphicx}%
\usepackage{multirow}%
\usepackage{amsmath,amssymb,amsfonts}%
\usepackage{amsthm}%
\usepackage{mathrsfs}%
\usepackage[title]{appendix}%
\usepackage{xcolor}%
\usepackage{textcomp}%
\usepackage{manyfoot}%
\usepackage{booktabs}%
\usepackage{algorithm}%
\usepackage{algorithmicx}%
\usepackage{algpseudocode}%
\usepackage{listings}%
\usepackage{array}

\theoremstyle{thmstyleone}%
\theoremstyle{thmstyletwo}%

\theoremstyle{thmstylethree}%

\begin{document}

\title[Article Title]{ InSight-R: A Framework for Risk-informed Human Failure Event Identification and Interface-Induced Risk Assessment Driven by AutoGraph}

\author[1,2]{ \sur{Xingyu Xiao}}\email{xxy23@mails.tsinghua.edu.cn}
\author[1,2]{\sur{Jiejuan Tong}}
\author[3]{\sur{Peng Chen}}
\author[1]{ \sur{Jun Sun}}
\author[1]{ \sur{Zhe Sui}}
\author*[1]{\sur{Jingang Liang}}\email{jingang@tsinghua.edu.cn}
\author[1,2]{ \sur{Hongru Zhao}}
\author[1,2]{ \sur{Jun Zhao}}
\author[1]{\sur{Haitao Wang}}

\affil[1]{\orgdiv{Institute of Nuclear and New Energy Technology}, \orgname{Tsinghua University}, \orgaddress{ \city{Beijing}, \postcode{100084}, \country{China}}}       

\affil[2]{\orgdiv{National Key Laboratory of Human Factors Engineering}, \orgaddress{ \city{Beijing}, \postcode{100094}, \country{China}}}  

\affil[3]{\orgdiv{Software Institute}, \orgname{Chinese Academy of Sciences}, \orgaddress{ \city{Beijing}, \postcode{100086},  \country{China}}}

\abstract{Human reliability remains a critical concern in safety-critical domains such as nuclear power, where operational failures are often linked to human error. While conventional human reliability analysis (HRA) methods have been widely adopted, they rely heavily on expert judgment for identifying human failure events (HFEs) and assigning performance influencing factors (PIFs). This reliance introduces challenges related to reproducibility, subjectivity, and limited integration of interface-level data. In particular, current approaches lack the capacity to rigorously assess how human-machine interface design contributes to operator performance variability and error susceptibility. To address these limitations, this study proposes a framework for risk-informed human failure event identification and interface-induced risk assessment driven by AutoGraph (InSight-R). By linking empirical behavioral data to the interface-embedded knowledge graph (IE-KG) constructed by the automated graph-based execution framework (AutoGraph), the InSight-R framework enables automated HFE identification based on both error-prone and time-deviated operational paths. Furthermore, we discuss the relationship between designer-user conflicts and human error. The results demonstrate that InSight-R not only enhances the objectivity and interpretability of HFE identification but also provides a scalable pathway toward dynamic, real-time human reliability assessment in digitalized control environments. This framework offers actionable insights for interface design optimization and contributes to the advancement of mechanism-driven HRA methodologies.}


\maketitle

\section{Introduction}\label{introduction}

Human error remains a leading contributor to failures in complex socio-technical systems such as nuclear power plants, aviation, and healthcare, where safety-critical operations depend on accurate and timely human decisions \cite{reason1990human, hollnagel1998cognitive}. Human reliability analysis (HRA) methods have been widely used to model operator behavior and assess the likelihood of human failure events (HFEs) \cite{xiao2024krail}. However, prevailing HRA approaches are often constrained by their reliance on expert judgment, particularly in the identification of HFEs and the assignment of performance influencing factors (PIFs) \cite{xiao2024krail, xiao2025dynamic}.

In traditional HRA frameworks such as the integrated human event analysis system for event and condition assessment (IDHEAS-ECA), HFEs are primarily determined through expert elicitation, a process that, while practical, suffers from limited reproducibility, insufficient transparency, and weak theoretical grounding \cite{xiao2024emergency}. This expert-centric process makes it difficult to systematically account for contextual complexity, dynamic task variations, and operator interface interactions. Similarly, the quantification of PIFs, factors that shape human performance, such as task complexity, information quality, or time pressure, is also typically handled through subjective scoring mechanisms \cite{xiao2025cognitive}. These scores are often imprecise, lack formalized computational definitions, and are disconnected from objective system data.

Among these factors, the human-machine interface plays a particularly critical role in shaping operator behavior and influencing error occurrence. Yet, current methods seldom incorporate rigorous assessments of interface complexity or operator-interface interactions into PIF modeling. To improve the predictive power and practical utility of HRA, there is a growing need for frameworks that go beyond expert intuition and incorporate measurable, mechanism-based insights into operator behavior.

Additionally, in recent years, nuclear power plant (NPP) control rooms have been progressively transitioning toward advanced digital human–machine interfaces and computerized systems \cite{xiao2024emergency}. Some may argue that with the development of autonomous technologies, the future of nuclear operations lies in full automation or "unmanned" plants, thereby diminishing the importance of human factors. However, "unmanned" does not imply "human-error-free." So-called unmanned systems are underpinned by extensive human involvement, including design, deployment, configuration, maintenance, and remote intervention. No nuclear power plant is truly devoid of human presence; rather, human factors become less visible but remain deeply embedded within the system. Achieving effective automation thus depends not on eliminating human roles, but on better understanding and mitigating human error. If we cannot clearly articulate the mechanisms by which human errors occur, then critical design decisions, such as whether a system should assist or replace human operators, when and how automation should intervene, or how to foster operator trust, risk being reduced to ad hoc or intuition-driven engineering.

In response to these challenges, this study proposes a novel framework titled interface-embedded human error and risk assessment (IE-HERA). The objective of this framework is to integrate knowledge graph modeling with operational behavior data to enable automated risk-informed HFE identification and quantitative assessment of interface-induced risks. Specifically, the contributions of this paper are fourfold:

\begin{itemize}
    \item We propose InSight-R, a hybrid framework combining structured knowledge modeling with data-driven behavioral analysis;
    \item We develop an HFE recognition method based on interface-embedded knowledge graphs (IE-KG), capturing the semantics and structure of human-machine interactions;
    \item We construct a set of novel metrics to quantify cognitive conflict between users and interfaces, thereby enabling measurable estimation of PIFs;
\end{itemize}

This paper is structured as follows: Section \ref{Related Work} presents the related work, Section \ref{Methodology} outlines the methodology, Section \ref{Case Study} discusses the results and evaluation of the case study, and Section \ref{Conclusion and Discussion} concludes with a discussion.

\section{Background and Related Work}\label{Related Work}

\subsection{Human Failure Events (HFE) in HRA}
Human reliability analysis (HRA) has emerged as a critical discipline for understanding and managing risks in complex socio-technical systems \cite{park2020inter}. At the core of any HRA lies the concept of the human failure event (HFE). HFEs are defined as the basic unit of analysis in HRA, representing specific human errors that can lead to adverse consequences for system safety. The field of HRA has progressed into its third generation \cite{xiao2025dynamic}, characterized by an increasing emphasis on dynamic task environments and cognitive theory-based modeling. A representative example of this evolution is the integrated human event analysis system for event and condition assessment (IDHEAS-ECA), developed by the U.S. Nuclear Regulatory Commission (NRC) \cite{xiao2024krail}. Rooted in the broader IDHEAS-General framework (IDHEAS-G) outlined in NUREG-2198, IDHEAS-ECA offers a structured eight-step methodology for analyzing human failure events (HFEs) to support risk-informed decision-making \cite{xing2020integrated}.

However, the determination of HFEs is heavily reliant on expert judgment. Such judgments are inherently subjective, shaped by individual backgrounds, mental models, and cognitive frameworks, making them vulnerable to inconsistency and bias. For complex procedures, such as those described in Section \ref{Case Study}, it is often difficult to distinguish critical tasks from routine actions solely based on procedural descriptions, posing a significant challenge for expert-based analysis.

In response to these limitations, this study proposes a knowledge-driven and data-enhanced framework designed to support real-time HRA in main control room environments. Specifically, by leveraging the interface element knowledge graph (IE-KG) and the HTRPM Tracker, a custom-developed tool capable of automatically collecting operator behavior data, this approach enables the objective identification of HFEs during training phases. In contrast to traditional methods that focus exclusively on error occurrence, our framework also captures “tail-end risks,” i.e., operation steps with significantly extended durations that may increase the likelihood of human error. As such, the proposed system offers a scientifically grounded pathway toward real-time, data-informed HRA applicable in digitalized control environments.

\subsection{Performance Influencing Factors (PIFs) Related to Interface Complexity}

Human error is commonly defined as an unintentional, inappropriate, or untimely action, inaction, or decision that deviates from established standards, procedures, or expectations, and may lead to undesirable outcomes \cite{reason1990human}. Importantly, human error is often a symptom of broader systemic issues rather than merely a reflection of individual incompetence. Poorly designed or overly complex systems are significant external contributors to error occurrence \cite{thimbleby2015unreliable}. In particular, complex interfaces, characterized by a high density of elements, disorganized layouts, or poorly structured information, require users to expend substantial cognitive resources (e.g., attention, working memory) \cite{roll2019human} to perceive, interpret, filter information, and make decisions. Moreover, human cognitive capacity is inherently limited, especially under conditions of time pressure or elevated stress. When the cognitive demands imposed by the interface exceed an individual's available mental resources, the efficiency and accuracy of information processing deteriorate, thereby increasing the likelihood of error. Such cognitive overload may manifest as slips due to attentional lapses, forgetfulness caused by working memory saturation, or judgment errors stemming from information overload and integration difficulties. Therefore, to meaningfully understand and quantify the impact of complexity on human error, it is essential not only to measure the structural complexity of the interface itself but also to assess its actual effect on user cognitive load, and, crucially, to link fluctuations in cognitive load to observable error behaviors. Merely counting the number of interface elements may be insufficient to capture the full extent of cognitive strain induced by specific element arrangements or interaction logic.

However, existing HRA methodologies rely heavily on expert judgment for the evaluation of human-system interfaces (HSIs) \cite{o2020human}. Experts typically perform coarse-grained classifications based on personal experience, resulting in assessments that are inherently subjective and difficult to replicate. Most notably, these methods lack real-time applicability and primarily focus on the occurrence of human errors, rather than on identifying and addressing error precursors. Although frameworks such as IDHEAS-ECA incorporate interface factors via performance influencing factors (PIFs), the criteria used to classify interface-related influences remain vague, qualitative in nature, and devoid of standardized, quantitative metrics. The absence of such quantifiable indicators, combined with differing interpretations among analysts, poses significant challenges to consistent and objective expert evaluation. Moreover, for several macro-cognitive processes, including detection (D), understanding (U), decision making (DM), action execution (E), and inter-team coordination (T), data availability remains particularly limited \cite{xing2020integrated}.

Therefore, establishing a scientifically grounded approach for quantifying interface complexity and clarifying its relationship with human error has become a critical challenge in both human factors engineering and human-computer interaction research. Relying solely on designers' intuition or subjective experience is no longer sufficient; instead, there is an urgent need to develop empirical, data-driven assessment methods and models that can support objective, repeatable evaluations of interface-induced risks.

\subsection{Interface Design and Human Error}

The design of user interfaces is intrinsically linked to human performance, acting as a critical determinant of both efficiency and fallibility. Effective user interface design is not merely a technological endeavor but one that deeply involves the study of human cognition, perception, and behavior \cite{stone2005user}. The overarching goal is to create interfaces that are not only easy to use and aesthetically pleasing but also enhance operator efficiency, productivity, and overall system effectiveness \cite{thimbleby1990user}.  Conversely, poorly designed interfaces can lead to significant user frustration, impede task performance, and, most critically, become direct precursors to costly and sometimes catastrophic human errors \cite{morland1983human}.  Empirical evidence supports this, with studies indicating that a substantial percentage of unplanned industrial shutdowns and process upsets can be traced back to deficiencies in HMI design \cite{carlgren2014design}.

In recent years, nuclear power plant (NPP) control rooms have been progressively transitioning toward advanced digital human–machine interfaces and computerized systems. While digitalization offers significant improvements in automation and information accessibility, it also introduces new challenges related to software reliability and interface complexity. If human factors are not adequately considered during the design process, digital interfaces may give rise to novel types of human–machine interaction errors and require operators to adapt to unfamiliar interaction paradigms. In safety-critical systems such as NPPs, poor interface design can lead to human errors with potentially severe consequences.

Some may argue that with the development of autonomous technologies, the future of nuclear operations lies in full automation or "unmanned" plants, thereby diminishing the importance of human factors. However, "unmanned" does not imply "human-error-free." So-called unmanned systems are, in fact, underpinned by extensive human involvement—including design, deployment, configuration, maintenance, and remote intervention. No nuclear power plant is truly devoid of human presence; rather, human factors become less visible but remain deeply embedded within the system. Achieving effective automation thus depends not on eliminating human roles, but on better understanding and mitigating human error.

If we cannot clearly articulate the mechanisms by which human errors occur, then critical design decisions, such as whether a system should assist or replace human operators, when and how automation should intervene, or how to foster operator trust, risk being reduced to ad hoc or intuition-driven engineering.

\section{ Methodology}\label{Methodology}

\subsection{Overview of InSight-R Framework}

\begin{figure}[h]
\centering
\includegraphics[width=1.0 \textwidth]{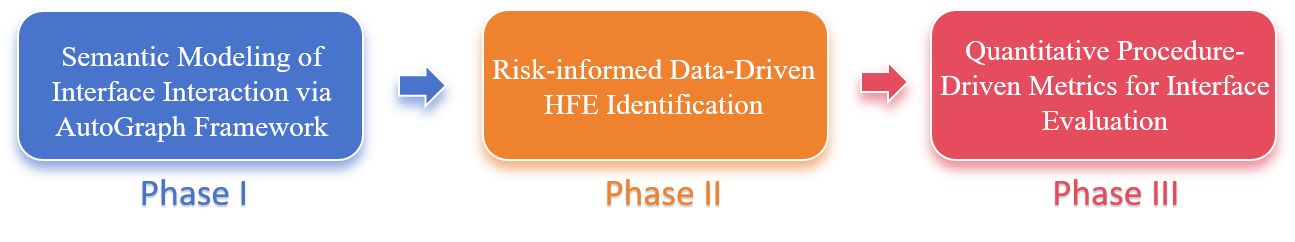}
\caption{Overview of the Insight-R Framework}\label{workflow}
\end{figure}

As discussed in Section~\ref{introduction}, current approaches to human failure event (HFE) identification heavily rely on expert judgment, which limits their objectivity and reproducibility. Additionally, these methods often lack a rigorous mechanism for addressing error-of-commission (EOC) scenarios, and existing evaluations of human–system interfaces are similarly dependent on subjective expert knowledge. To address these limitations, we propose the Insight-R framework, a risk-informed approach for HFE identification and interface-induced risk assessment, powered by the AutoGraph framework. The overall structure of Insight-R is illustrated in Figure~\ref{workflow}, comprising three key phases: Phase I, semantic modeling of interface interaction using AutoGraph; Phase II, risk-informed, data-driven HFE identification; and Phase III, quantitative evaluation of interface quality based on objective metrics. Each of these phases is detailed in the following sections.

\subsection{Phase I: Semantic Modeling of Interface Interaction via AutoGraph Framework}

\begin{figure}[h]
\centering
\includegraphics[width=0.9\textwidth]{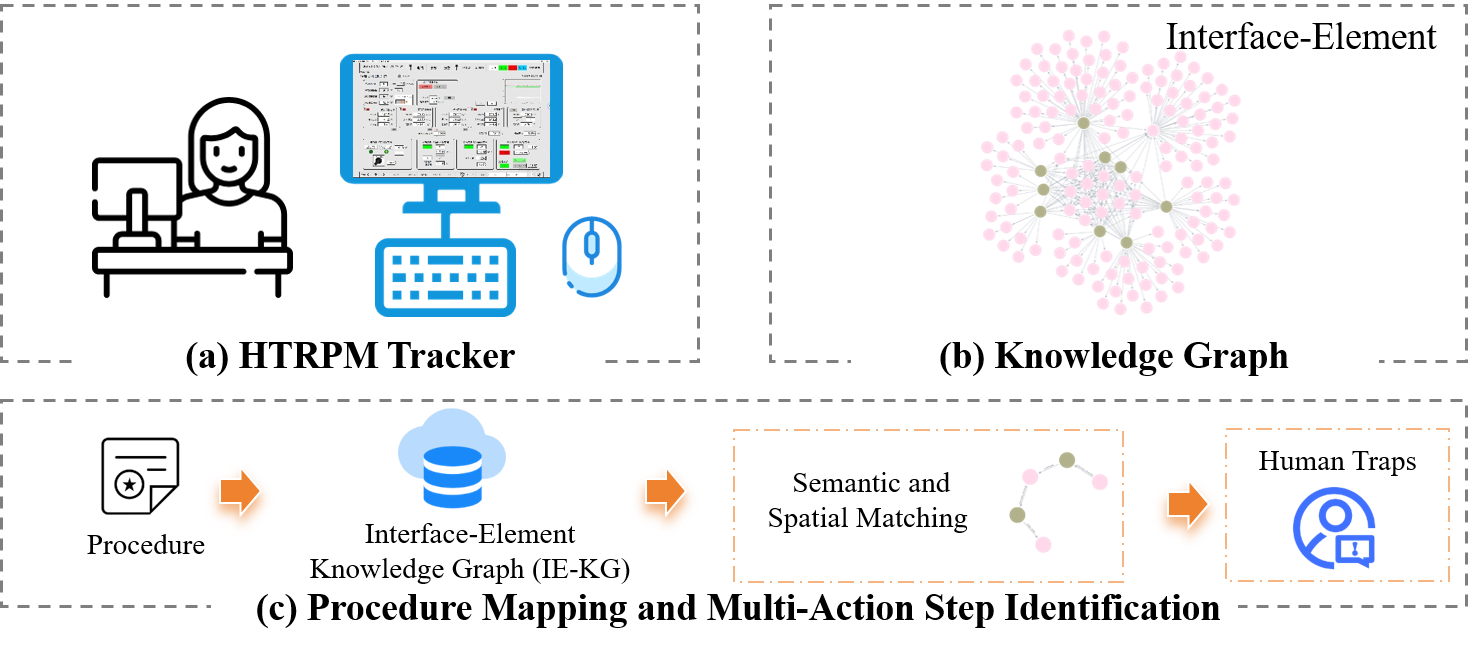}
\caption{Semantic Modeling of Interface Interaction via AutoGraph Framework}\label{AutoGraph}
\end{figure}

The AutoGraph framework establishes a structured and extensible foundation for semantic modeling and automation of interface interactions in digitalized nuclear control rooms. In this section, we introduced Parts I-III of the AutoGraph framework, as shown in Figure \ref{AutoGraph}, which enable the mapping of procedural tasks to the interface knowledge graph and the generation of corresponding execution paths.

In Phase I, a customized tracker was used to unobtrusively capture operator actions in the HTRPM simulation environment, including input events, spatial coordinates, timestamps, and contextual screenshots. In Phase II, this data was used to construct an Interface-Embedded Knowledge Graph (IE-KG) that formalizes GUI structure, with nodes representing interface elements and edges capturing hierarchical or logical relations. In Phase III, procedural steps were mapped to execution paths within the IE-KG. Paths requiring multiple sequential actions were flagged as multi-action steps, allowing for task complexity analysis and identification of potentially error-prone segments. Together, these components define the core structure and methodology of the IE-KG, comprising interface elements, spatial coordinates, operation mappings, and inter-element logical associations. This unified representation enables scalable, data-informed modeling of human-system interaction and lays the groundwork for subsequent HRA modules, such as error risk estimation, performance bottleneck analysis, and dynamic interface evaluation.

\subsection{Phase II: Risk-informed Data-Driven HFE Identification}

To enable data-driven identification of human failure events (HFEs), we first collected empirical data from human-in-the-loop experiments simulating representative scenarios in a digital nuclear power plant control environment. The dataset comprises four key components: (1) operator action sequences, capturing the ordered execution of interface interactions; (2) cursor trajectory data, recording the spatial movements across the interface; (3) error annotations, labeling erroneous operations based on real-world experiments; and (4) task duration records, indicating the time consumed in the human-in-the-loop experiments for each operational step. These heterogeneous data sources were systematically integrated and mapped onto the interface-element knowledge graph (IE-KG) constructed through the AutoGraph framework. Each node in the IE-KG represents a unique interactive element (e.g., a button) on the interface, enriched with spatial, semantic, and procedural attributes. By aligning the empirical data with the corresponding nodes, we enabled traceable analysis of operator behavior at the element level.

To uncover latent risks in operational behavior, we introduced two types of risk-index pathway identification: (1) error paths, defined as frequently traversed action sequences with a high incidence of operational errors. These paths highlight procedural segments where human reliability is repeatedly compromised; (2) time paths, defined as action sequences exhibiting significant deviation in task duration, particularly those that fall in the upper tail of the time distribution. These paths are indicative of increased cognitive workload, interface complexity, or situational uncertainty. By analyzing these risk-index paths, we identified safety-critical interface nodes associated with either high error rates or temporal anomalies. These nodes serve as objective indicators of potential risk-informed human performance vulnerabilities.

\begin{figure}[h]
\centering
\includegraphics[width=0.7\textwidth]{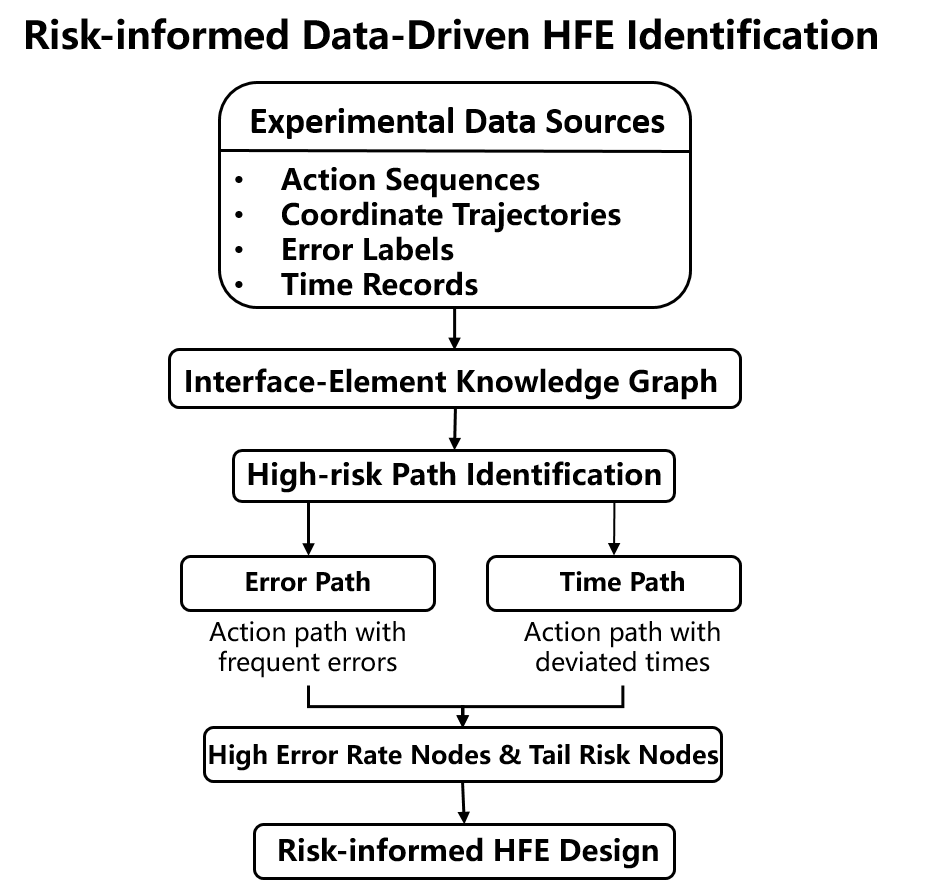}
\caption{The workflow of risk-informed data-driven HFE identification}\label{HFE_identification}
\end{figure}

Subsequently, we linked the identified high-risk nodes to the specific procedures and steps in which they appear. If a given procedural step involves one or more high-risk nodes, it is labeled as a candidate risk-informed human failure event (HFE). Furthermore, procedures that concentrate a large number of such nodes across multiple steps were prioritized for in-depth HFE analysis. This approach allows for a more objective and scalable identification of HFEs, in contrast to traditional expert-driven methods, and supports the development of data-informed reliability enhancement strategies. The complete workflow is illustrated in Figure \ref{HFE_identification}.

\subsection{Phase III: Quantitative Procedure-Driven Metrics for Interface Evaluation}\label{Quantitative Metrics for Interface Evaluation}

Current evaluations of the impact of human-system interfaces (HSIs) on human error predominantly rely on subjective judgment. For instance, while the IDHEAS-ECA framework includes interface-related performance influencing factors (PIFs), its classification criteria remain ambiguous, primarily qualitative, and lacking in standardized quantitative measures. Moreover, data availability for the five macro-cognitive processes, detection (D), understanding (U), decision making (DM), execution (E), and inter-team coordination (T), is limited, with most processes being sparsely supported by empirical data \cite{xing2020integrated}.

In phase I, a pre-constructed interface-element knowledge graph (IE-KG) serves as a structured digital representation of the interface, effectively mapping the PMSim environment into a network format referred AutoGraph. The IE-KG encodes rich metadata, including element names, coordinates, and hierarchical relationships, allowing for data-driven analysis of interface structure. Building upon the IE-KG, we propose a set of interface-user conflict metrics to characterize potential sources of cognitive and operational strain. These metrics include: visual density (VD), semantic interference density (SID), and interaction span (IS).

\textbf{Visual density (VD)} is defined as the ratio of the total area occupied by visible interface elements to the overall area of the interface:

\begin{equation}
VD = \frac{N_{target}}{N_{elements}}
\end{equation}
where $N_{target}$ denotes the number of the target element, which is always denoted as one; $N_{elements}$ represents the total number of all visible elements on the interface.

\textbf{Semantic interference density (SID)} quantifies the potential for semantic confusion among interface elements. If the name of a target element (e.g., "Main Pump Start") is highly similar to the names of several non-target elements on the same interface (e.g., "Main Pump Shutdown", "Auxiliary Pump Start"), users may experience increased cognitive load and a higher risk of misoperation during recognition and interaction.

Let the name of the target parameter be denoted as $T$, and the set of all other visible parameter names on the current interface be represented as $PN = {PN_{1}, PN_{2}, ..., PN_{n}}$. To quantify potential semantic interference, we compute the cosine similarity between $T$ and each $PN_{i}$, using vector representations derived from the text-embedding-ada-002 model. The resulting similarity scores are denoted as $Sim_{i} = similarity(T, PN_{i})$. A similarity threshold is applied to identify potential interference; in this study, we consider $Sim_{i} > 0.8$ as indicative of semantic interference.

To assess SID across an entire interface, we compute the number of element pairs whose name similarity exceeds a predefined threshold in Equation \ref{SID}, thereby identifying clusters of semantically confusable items.

\begin{equation}\label{SID}
    SID=\frac{N_{\text{high-similarity}}}{N_{\text{{total}}}}
\end{equation}
where $N_{high-similarity}$ denotes the number of element pairs on the interface whose name similarity exceeds a predefined threshold, and $N_{total}$ represents the total number of possible element pairs on the interface.

\textbf{Interaction span (IS)}. This metric quantifies the spatial distance that must be traversed, by mouse movement or visual attention, in order to complete a specific operation, thereby reflecting the operational workload associated with physical or cognitive effort. It is defined as:
\begin{equation}
    IS = \frac{\text{Euclidean Distance}(P_i, P_{i+1})}{Distance_{longest}}
\end{equation}
where $P_i$ and $P_{i+1}$ are consecutive points in the recorded interaction trajectory. The variable $Distance_{longest}$ represents the maximum possible cursor travel distance across the interface, reflecting the longest spatial extent over which the mouse can be moved within the interface layout.

\section{Case Study and Experimental Results}\label{Case Study}

\subsection{ Scenario Description}\label{Scenario Description}

Five task scenarios were designed for experimental validation, each comprising 4 to 7 procedural sub-steps. Participants were instructed to execute these steps sequentially on the HTRPMSim digital simulator.

Six graduate students specializing in nuclear science and technology (two master's and four doctoral candidates) participated in the study. Prior to testing, all participants received comprehensive training to ensure full familiarity with the simulation environment. To emulate operational stress and induce error-prone behavior, a time-based incentive was implemented: the fastest participant across all tasks received a performance-based reward. Each session lasted approximately 20–40 minutes, yielding around three hours of recorded data. Three types of data were collected: (1) screen recordings capturing simulator operations; (2) real-time cursor trajectories and click events using a custom tool (tracker.exe); and (3) third-person video footage to document external operator behavior. 

To mirror actual operational protocols, the experiment incorporated a widely used human performance tool (HPT), the two-step verification card. Participants marked each intended action during procedure reading and confirmed execution with a corresponding checkmark. This tool, designed to mitigate execution of omission (EOO), proved effective: no EOO instances were observed throughout the experiment. Additionally, we constructed a partial IE-KG to represent selected HTGRSim interface panels. For full details on IE-KG construction.

\subsection{HFE Identification Results} \label{HFE Identification Results}

Traditionally, the identification of human failure events (HFEs) has relied heavily on expert judgment. However, with the introduction of our interface-embedded knowledge graph (IE-KG), this process is transformed into a data-driven approach that enables more objective and reproducible HFE identification. Specifically, HFEs are identified from two complementary perspectives: error-prone operational paths, which reflect frequent execution errors, and time-deviated operational paths, which indicate tasks with significantly prolonged durations and potential cognitive overload. This dual-perspective analysis supports risk-informed decision-making in human reliability assessment and interface optimization.

\subsubsection{Identification of Error-Prone Operational Paths}\label{Identification of Error-Prone Operational Paths}
\begin{figure}[h]
\centering
\includegraphics[width=0.9\textwidth]{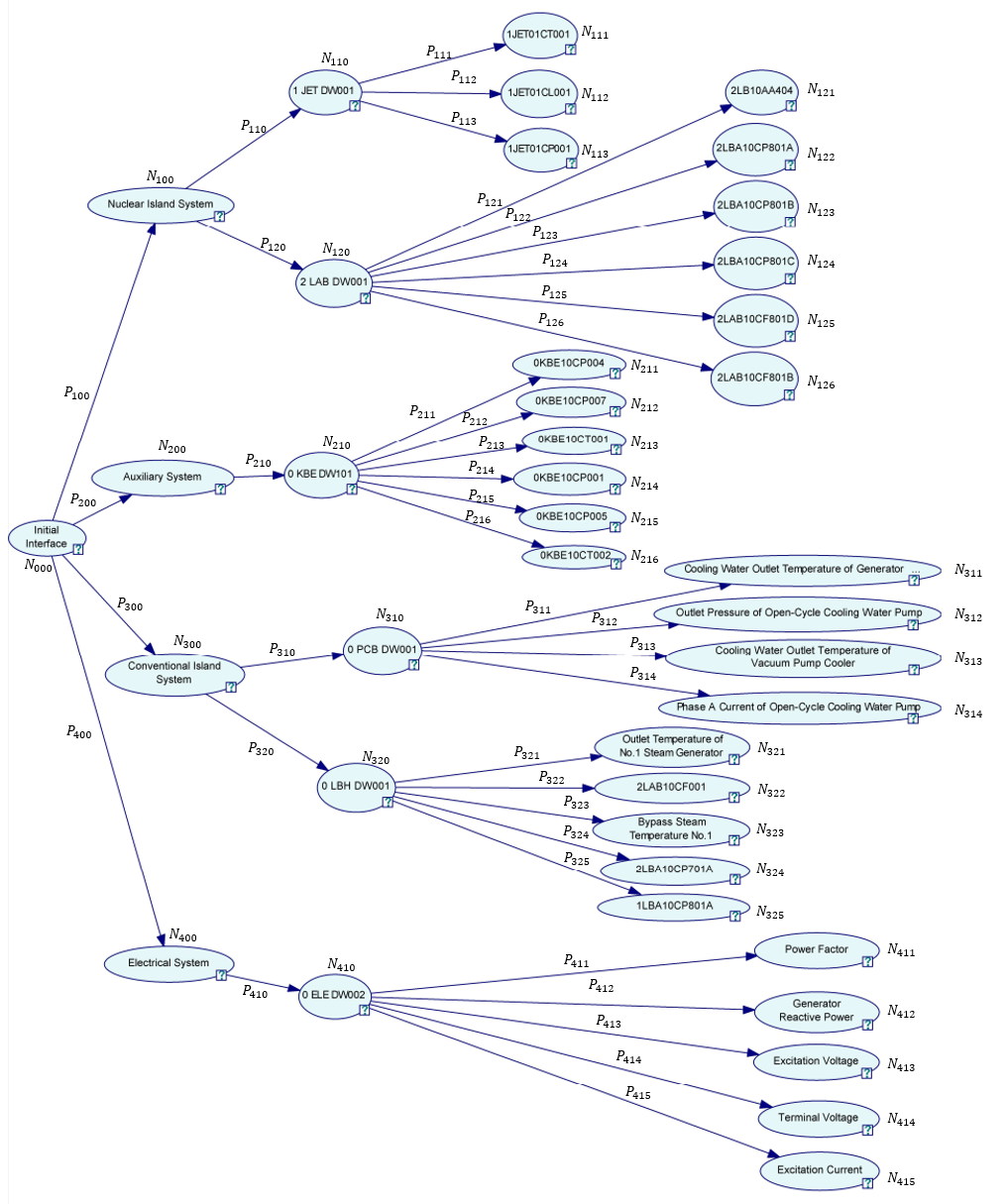}
\caption{Structure of the Bayesian Network Reflecting the Experimental Procedure}\label{genle}
\end{figure}

We mapped the collected error data onto a Bayesian network to facilitate various probabilistic analyses. The structure of the constructed Bayesian network is illustrated in Figure \ref{genle}. Each node in the network is assigned a unique identifier and labeled according to its hierarchical affiliation within the system. For example, node $N_{400}$ represents the electrical system, while node $N_{410}$ denotes a specific parameter within the electrical system, namely "0 ELEDW002." Nodes $N_{411}$ through $N_{415}$ correspond to the set of parameters under $N_{410}$, including power factor, generator reactive power, excitation voltage, terminal voltage, and excitation current. Network paths are denoted by $P$, where, for instance, the path from $N_{100}$ to $N_{110}$ is encoded as $P_{110}$.

\begin{figure}[h]
\centering
\includegraphics[width=1.0\textwidth]{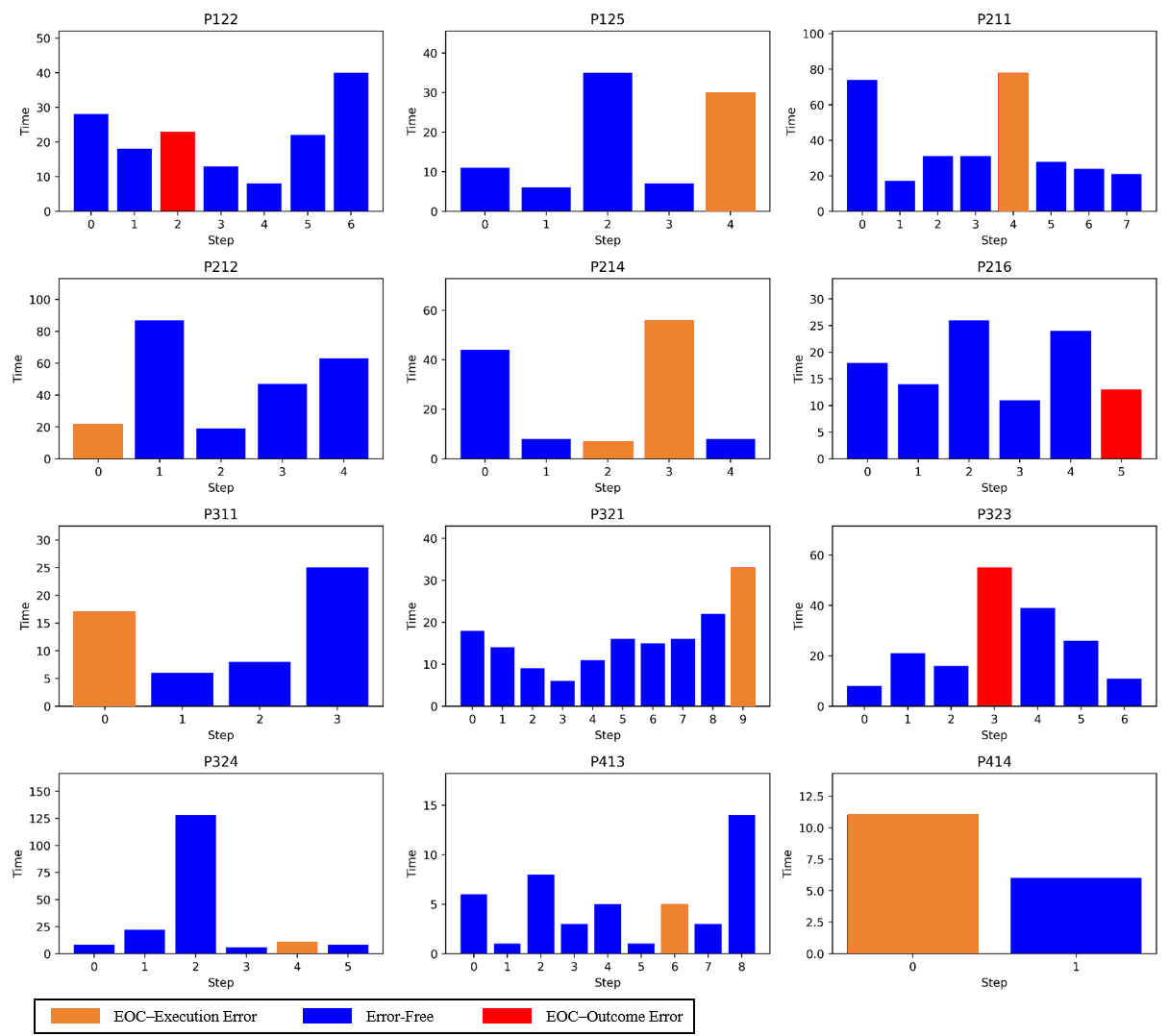}
\caption{Interface navigation path corresponding to the procedural task: “Check whether the parameter 2LBA10CP801C under Nuclear Island System 2 LAB DW001 equals 13.86 MPa.”}\label{error_time_}
\end{figure}

Among all the paths illustrated in Figure \ref{genle}, only the following paths: $P_{122}$, $P_{125}$, $P_{211}$, $P_{212}$, $P_{214}$, $P_{216}$, $P_{311}$, $P_{321}$, $P_{323}$, $P_{324}$, $P_{413}$, and $P_{414}$, were associated with human errors. All of these instances were categorized as errors of commission (EOC). We visualized the distribution of operation times for the tasks associated with these error-prone paths based on experimental data, as shown in Figure \ref{error_time_}. The orange-highlighted paths represent execution errors, while the red-highlighted paths indicate outcome errors. As shown, outcome errors are relatively infrequent, occurring in only three paths: $P_{122}$, $P_{216}$, and $P_{123}$.

The results indicate that the occurrence of errors exhibits a certain degree of randomness. For example, some errors occurred on the longest-duration path (e.g., $P_{321}$), while others appeared on relatively short-duration paths (e.g., $P_{324}$). Nevertheless, a majority of the errors (7 out of 12) occurred on paths with relatively longer operation times. This observation provides preliminary empirical support for the hypothesis proposed in Section \ref{Detection of Time-Deviated Operational Paths}, which suggests that paths with longer durations may be more susceptible to errors and could be considered as residing within the "tail-end risk" region of task performance.

\subsubsection{Detection of Time-Deviated Operational Paths}\label{Detection of Time-Deviated Operational Paths}

To compare the performance of different operational paths, we conducted a time-based risk analysis focused on identifying both stable paths and those associated with elevated tail-end timing risks, defined as the probability that a path’s task duration falls within the longer range compared to other paths of the same type. Prior studies have demonstrated a strong correlation between extended task durations and increased cognitive workload \cite{chen2024influence}, which in turn elevates the likelihood of human error. In this work, we adopt the assumption that longer execution time implies a higher probability of performance degradation or failure.

To model the distribution of operator-required time for each path, we follow the technical guidance provided by the IDHEAS-ECA methodology \cite{chang2021idheas}, which recommends a lognormal distribution for human task durations in control room settings. The distribution’s shape parameter $\sigma$ is fixed at 0.28, an empirically validated constant for nuclear operational contexts. The scale parameter $\mu$ is calculated as follows: If the median task completion time is known, $\mu$ is computed directly as the natural logarithm of the median, i.e., $\mu$ = ln(median); If only the 95th percentile time ($t_{95}$) is available, the median is approximated using the relation \ref{median}, as suggested in the IDHEAS-ECA technical basis.

\begin{equation} \label{median}
median\approx t_{95}/ 1.585
\end{equation}

Subsequently, the operator-required time for each path is modeled as:

\begin{equation}
T_{reqd}\sim  Lognormal(\mu ,\sigma=0.28)
\end{equation}

This lognormal formulation enables probabilistic reasoning about operator timing performance and supports scenario-level human reliability assessment under time constraints.

\begin{figure}[h]
\centering
\includegraphics[width=1.0\textwidth]{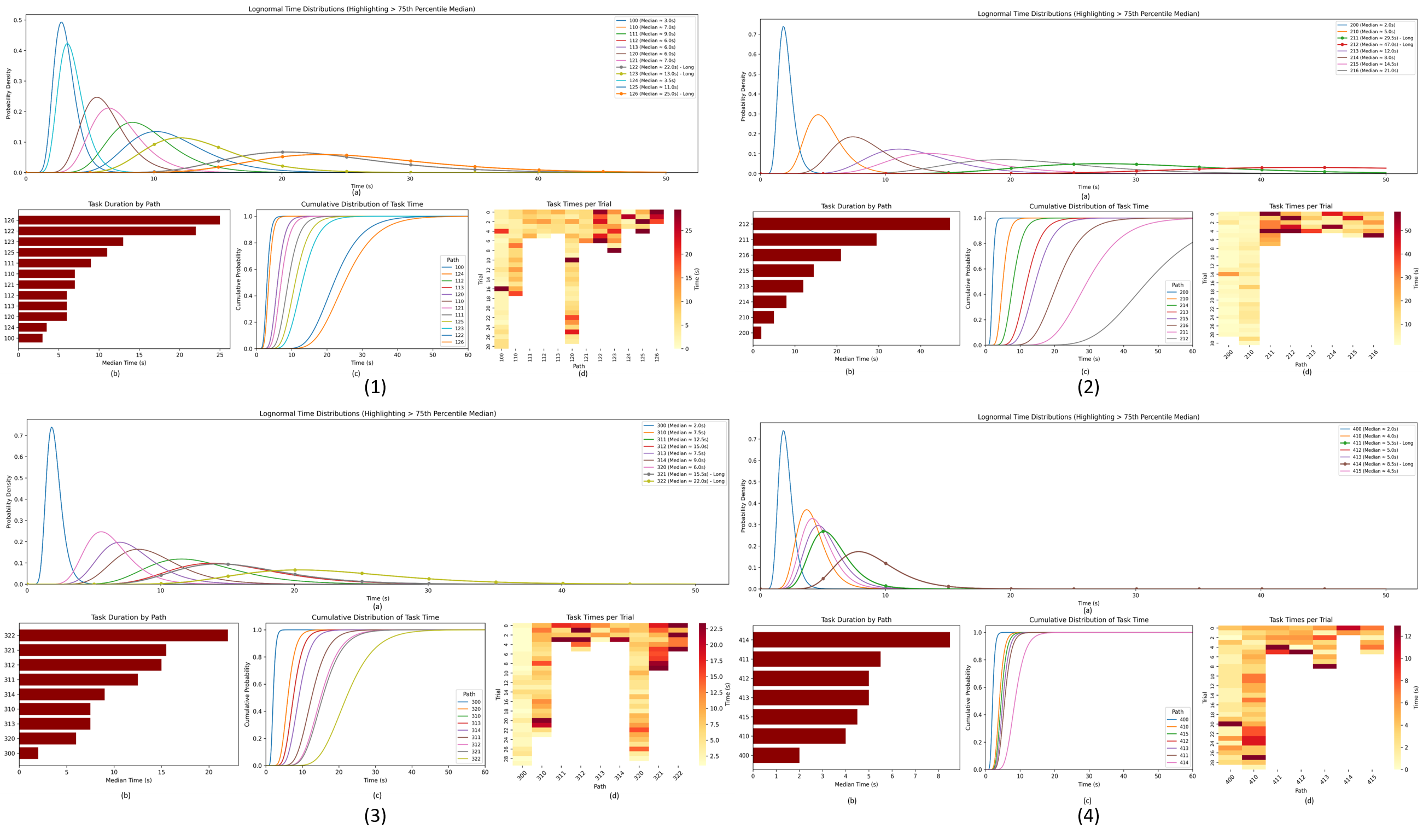}
\caption{Distribution of Human Task Duration Across Operational Paths by System Category}\label{time}
\end{figure}

As illustrated in Figure~\ref{time}, the experimental results reveal that certain paths, specifically, $P_{211}$, $P_{212}$ (nuclear island system), $P_{122}$, $P_{123}$,$ P_{126}$ (auxiliary system), $P_{321}$, $P_{322}$ (conventional island system), and $P_{411}$, $P_{414}$ (electrical system), exhibited significantly longer task durations compared to other paths within the same system category. These paths are thus identified as potential tail-risk candidates, warranting further scrutiny in reliability evaluation and procedural design.

By synthesizing the results from both error-prone operational paths and time-deviated operational paths, we identified a set of risk-informed Human Failure Events (HFEs). These HFEs reflect operational steps that are either frequently associated with execution errors or exhibit significantly prolonged task durations—both of which imply elevated cognitive demand and increased potential for human error. Based on the experimental procedures, the final list of identified HFEs includes:

\begin{itemize}
    \item $P_{122}$, $P_{323}$,$P_{216}$
    \item $P_{125}$, $P_{211}$, $P_{212}$, $P_{214}$, $P_{216}$, $P_{311}$, $P_{321}$, $P_{323}$, $P_{324}$, $P_{413}$, and $P_{414}$
    \item $P_{211}$, $P_{212}$ , $P_{122}$, $P_{123}$,$ P_{126}$, $P_{321}$, $P_{322}$ , $P_{411}$ and  $P_{414}$ 
\end{itemize}

These results provide a data-driven foundation for targeted HFE analysis and interface improvement strategies in digital nuclear control environments.

\subsection{Quantitative Procedure-Driven PIF Evaluation}

\begin{figure}[h]
\centering
\includegraphics[width=1.0\textwidth]{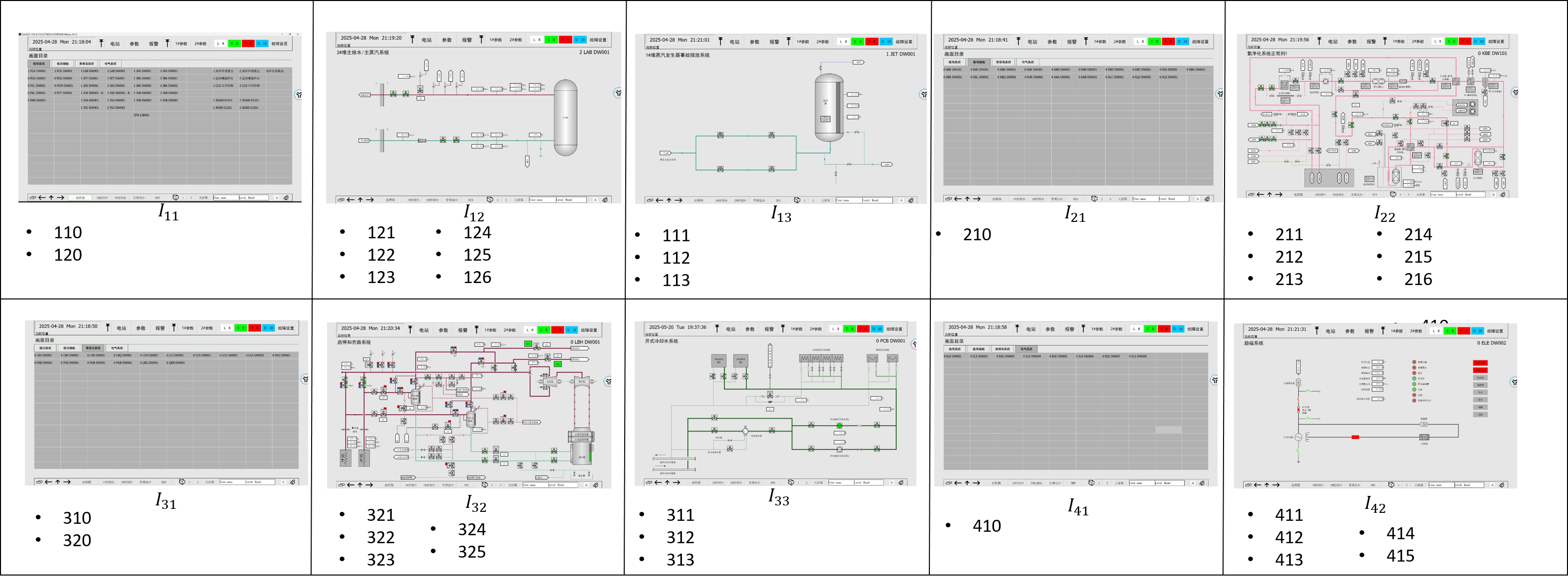}
\caption{Categorization and Examples of Flowchart-Based and Table-Based Interfaces in PMSim}\label{interface}
\end{figure}

We extracted and encoded the interface information involved in Section \ref{Scenario Description}, as illustrated in the example shown in Figure \ref{interface}. Based on the HTRPMSim simulation platform, the interfaces were categorized into two types: flowchart-based and table-based. The flowchart-based interfaces include $I_{32}$, $I_{33}$,$I_{42}$, $I_{12}$, $I_{22}$, and $I_{13}$, whereas the table-based interfaces comprise $I_{31}$, $I_{41}$, $I_{11}$, and $I_{21}$. As shown in the results of Section~\ref{Identification of Error-Prone Operational Paths}, errors are predominantly concentrated within the flowchart-based interface.

Currently, many existing approaches for interface evaluation overlook the content and objectives of operating procedures. In contrast, the evaluation method introduced in Section~\ref{Quantitative Metrics for Interface Evaluation} integrates both interface characteristics and the procedural goals, enabling a more comprehensive and context-aware assessment. To capture semantic similarity between procedural steps and interface elements, we employ the text-embedding-ada-002 model \cite{nussbaum2024nomic}.

\begin{figure}[h]
\centering
\includegraphics[width=1.0\textwidth]{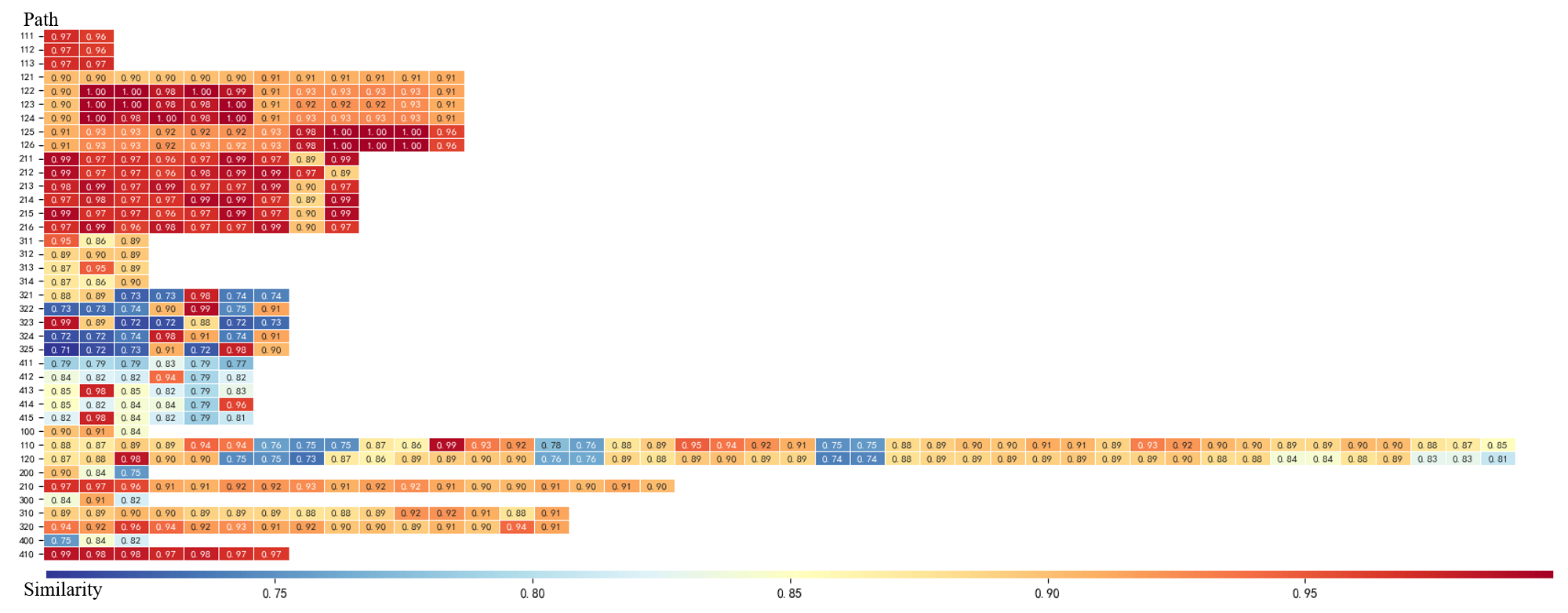}
\caption{Semantic Similarity Matrix for Parameter Names Indicating High Interference Risk Clusters}\label{similarity}
\end{figure}

A high semantic interference density (SID) indicates the presence of multiple parameters on the interface that are semantically similar to the target, thereby increasing the likelihood of misjudgment or operational errors. The computed semantic similarities between different parameter names are presented in Figure~\ref{similarity}. As shown, parameter groups such as $P_{111}$-$P_{113}$, $P_{122}$-$P_{126}$, $P_{211}$-$P_{216}$, and $P_{410}$ exhibit high SID values, suggesting elevated interface-induced risk. The results of the interface evaluation based on the proposed procedure-driven approach are summarized in Table~\ref{metric}.

\begin{table}[h]
\centering
\renewcommand{\arraystretch}{1.3} 
\begin{tabular}{>{\raggedright\arraybackslash}p{1.5cm} >{\raggedright\arraybackslash}p{1.5cm}>{\raggedright\arraybackslash}p{1.5cm}>{\raggedright\arraybackslash}p{2.5cm}>{\raggedright\arraybackslash}p{2.5cm}} 
\hline
\textbf{Path} & \textbf{VD}  & \textbf{SID} & \textbf{IS} & \textbf{PIF}
\\ 
\hline
$P_{100}$ & 1/4 & 0/3 & 1267.22/2654.05 & HSI0\\
$P_{110}$ & 1/43 & 1/42 & 511.59/2654.05 & HSI1\\
$P_{111}$ & 1/4 & 2/3 & 1280.07/2654.05 & HSI0 \\
$P_{112}$  & 1/4 & 2/3 & 1256.33/2654.05 & HSI0 \\
$P_{113}$ & 1/4 & 2/3 & 1243.96/2654.05 & HSI0 \\
$P_{120}$ & 1/43 & 1/42 & 735.97/2654.05 & HSI1\\
$P_{121}$ & 1/13 & 0/12 & 252.68/2654.05 & HSI1\\
$P_{122}$ & 1/13 & 5/12 & 913.84/2654.05 & HSI1\\
$P_{123}$ & 1/13 & 5/12 & 750.67/2654.05 &HSI1 \\
$P_{124}$ & 1/13 & 5/12 & 590.25/2654.05 & HSI1\\
$P_{125}$ & 1/13 & 5/12 & 786.33/2654.05 & HSI1\\
$P_{126}$ & 1/13 & 5/12 & 882.77/2654.05 & HSI1\\
\hline
$P_{200}$ & 1/4 & 0/3 & 1222.86/2654.05 & HSI0\\
$P_{210}$ & 1/19& 3/18 & 209.35/2654.05 & HSI1\\
$P_{211}$ & 1/20 & 8/19 & 1705.02/2654.05 &HSI5 \\
$P_{212}$ & 1/20 & 8/19 & 2220.46/2654.05 &HSI5 \\
$P_{213}$ & 1/20 & 8/19 & 588.80/2654.05 &HSI5\\
$P_{214}$ & 1/20 & 8/19 & 232.52/2654.05 & HSI5\\
$P_{215}$ & 1/20 & 8/19 & 1549.82/2654.05 & HSI5\\
$P_{216}$ & 1/20 & 8/19 & 657.01/2654.05 & HSI5 \\

\hline
$P_{300}$ & 1/4 & 0/3 & 1213.48/2654.05 &  HSI0\\
$P_{310}$ & 1/4 & 2/3 & 215.60/2654.05 & HSI1\\
$P_{311}$ & 1/5 & 0/4 & 1780.31/2654.05 & HSI0\\
$P_{312}$ & 1/5 & 0/4 & 1876.91/2654.05 & HSI0\\
$P_{313}$ & 1/5 & 0/4 & 1122.47/2654.05 & HSI0\\
$P_{314}$ & 1/5 & 0/4 & 1552.57/2654.05 & HSI0\\
$P_{320}$ & 1/14 & 1/13 & 186.10/2654.05 & HSI1\\
$P_{321}$ & 1/18 & 1/17 & 761.20/2654.05 & HSI5\\
$P_{322}$ & 1/18 & 1/17 & 623.87/2654.05 &HSI5 \\
$P_{323}$ &1/18 & 1/17 & 924.93/2654.05 & HSI5\\
$P_{324}$ & 1/18 & 1/17 & 1379.73/2654.05 &HSI5 \\
$P_{325}$ & 1/18 & 1/17 & 794.43/2654.05 & HSI5\\
\hline
$P_{400}$ & 1/4 & 0/3 & 1243.78/2654.05 & HSI0\\
$P_{410}$ & 1/8 & 7/7 & 395.25/2654.05 & HSI1\\
$P_{411}$ & 1/7 & 0/6 & 898.57/2654.05 & HSI0 \\
$P_{412}$ & 1/7 & 0/6 & 887.85/2654.05 & HSI0\\
$P_{413}$ & 1/7 & 1/6 & 859.37/2654.05 &HSI0 \\
$P_{414}$ & 1/7 & 1/6 & 858.08/2654.05 & HSI0\\
$P_{415}$ & 1/7 & 1/6 & 862.92/2654.05 &HSI0 \\
\hline
\end{tabular}
\caption{Quantitative Evaluation Results of Interface Metrics for PIF Assessment}
\label{metric}
\end{table}

\section{Discussion}

\subsection{Comparable HFE Analysis}
Traditional HFE identification approaches largely depend on expert judgment. In alignment with industry practice, we consulted several domain experts to understand how HFEs are typically determined. According to NUREG-1792 \cite{chang2016human}, HFEs are often identified during simulator-based training sessions through structured interviews and post-scenario debriefings. Experts tend to focus on key operator behaviors, specifically, actions that cause critical changes in system status, which may in turn affect the overall safety state of the plant. The experimental scenarios we designed follow this logic, focusing on parameter-search tasks that serve as precursors to such key actions.

During post-experiment interviews, participants consistently reported that tasks involving interface panels $I_{32}$ and $I_{22}$ were more cognitively demanding. Based on expert assessment alone, HFEs would likely be associated with systems 0LBH DW001 and 0KBE DW101, corresponding to the following procedure paths: $P_{321}$-$P_{325}$ and $P_{211}$-$P_{216}$.

\begin{figure}[h]
\centering
\includegraphics[width=0.7\textwidth]{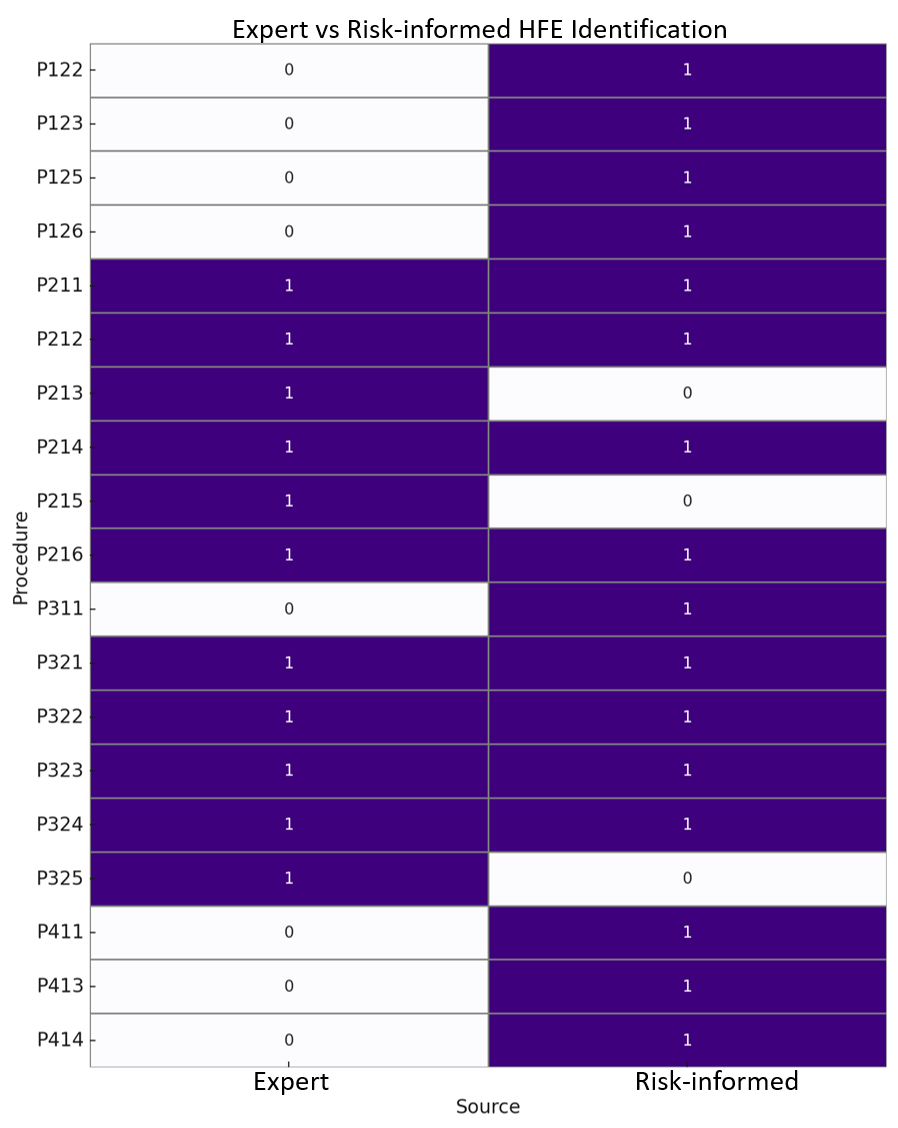}
\caption{Comparison of Expert-Based and Risk-Informed Identification of Human Failure Events}\label{output}
\end{figure}

In contrast, our risk-informed, data-driven HFE identification module produced the following HFE list in Section \ref{HFE Identification Results}. The comparison between expert-based and data-driven HFE identification is visualized in Figure~\ref{output}. As shown, expert-based methods primarily reflect subjective intuition and professional judgment. However, human behavior is inherently uncertain—operators may occasionally make errors even during seemingly simple tasks. Such instances often fall outside the scope of expert-derived HFEs, which tend to overlook low-complexity but high-variability actions. In contrast, the proposed risk-informed HFE identification framework exhibits a dynamic capability, allowing for real-time updates as new errors emerge. The HFE list can be incrementally refined by continuously incorporating historical performance data and feedback from operational scenarios. This adaptability makes the framework more representative of real-world operator behavior and more responsive to evolving risk patterns, offering a more robust and evidence-based foundation for human reliability analysis.

It is noteworthy that, in the experimental process, the participants involved were university students rather than licensed control room operators. As a result, the observed human failure events (HFEs) may exhibit greater variability and include error types that are less likely to occur in real-world scenarios. However, since the proposed framework is inherently dynamic and capable of continual updating with new data, this limitation does not compromise its applicability to actual operational contexts.

\subsection{Synergistic Integration with Traditional HRA Methods such as IDHEAS-ECA} \label{Integration with Traditional HRA Methods}

Moreover, our proposed method enables the quantification of an interface-related evaluation metric, providing a systematic and objective means to assess interface design characteristics. IDHEAS-ECA, developed by the U.S. Nuclear Regulatory Commission (NRC) \cite{chang2021idheas}, is currently the most authoritative human reliability analysis (HRA) methodology. It is grounded in five macro-cognitive theories and supported by validated empirical data. The range of performance influencing factors (PIFs) incorporated within the framework is comprehensive and well-established. Building upon this foundation, we adopted the IDHEAS-ECA structure to support the analysis of PIFs.

Within the IDHEAS-ECA framework, interface-related factors are primarily characterized using qualitative descriptions, as exemplified in Table~\ref{HSI_PIF}. In this study, we aim to map the interface evaluation metrics developed through our approach to the corresponding performance influencing factors (PIFs) defined in IDHEAS-ECA. This mapping enables the quantitative estimation of PIFs, thereby enhancing the objectivity and reproducibility of human reliability analysis. The PIFs confirmed through expert judgment for validation purposes are presented in Table~\ref{metric}.

\begin{table}[htbp]
\centering
\caption{PIF Weights for Human-System Interface in ref. \cite{xing2020integrated}} \label{HSI_PIF}
\begin{tabular}{p{1cm} p{6cm} c c c c c}
\toprule
\textbf{PIF} &\textbf{Attribute} & \textbf{D} & \textbf{U} & \textbf{DM} & \textbf{E} & \textbf{T} \\
\midrule
HSI0 &No impact – well designed HSI supporting the task & 1 & 1 & 1 & 1 & 1 \\
HSI1& Indicator is similar to other sources of information nearby & 1.5 & NA & NA & NA & NA \\
HSI2 &No sign or indication of technical difference from adjacent sources (meters, indicators) & 3 & NA & NA & NA & NA \\
HSI3& Related information for a task is spatially distributed, not organized, or cannot be accessed at the same time & 1.5 & 2 & NA & NA & NA \\
HSI4& Unintuitive or unconventional indications & 2 & NA & NA & NA & NA \\
HSI5& Poor salience of the target (indicators, alarms, alerts) out of the crowded background & 3 & NA & NA & NA & NA \\
HSI6& Inconsistent formats, units, symbols, or tables & 5 & NA & NA & NA & NA \\
HSI7& Inconsistent interpretation of displays & NA & 5.7 & NA & NA & NA \\
HSI8& Similarity in elements – Wrong element selected in operating a control element on a panel within reach and similar in design in control room & NA & NA & NA & 1.2 & NA \\
HSI9 & Poor functional localization – 2–5 displays/panels needed to execute a task & NA & NA & NA & 2 & NA \\
HSI10 & Ergonomic deficits: \newline
$\bullet$ Controls are difficult to maneuver \newline
$\bullet$ Labeling and signs of controls are not salient among crowd \newline
$\bullet$ Inadequate indications of states of controls - Small unclear labels, difficult reading scales \newline
$\bullet$ Maneuvers of controls are unintuitive or unconventional & NA & NA & NA & 3.38 & NA \\
HSI11 & Labels of the controls do not agree with document nomenclature, confusing labels & NA & NA & NA & 5 & NA \\
HSI12 & Controls do not have labels or indications & NA & NA & NA & 10 & NA \\
HSI13 & Controls provide inadequate or ambiguous feedback (i.e., lack of or inadequate confirmation of the action executed – incorrect, no information provided, measurement inaccuracies, delays) & NA & NA & NA & 4.5 & NA \\
HSI14 & Confusion in action maneuver states (e.g., automatic resetting without clear indication) & NA & NA & NA & 10 & NA \\
HSI15 & Unclear functional allocation (between human and automation) & NA & NA & NA & 9 & NA \\
\bottomrule
\end{tabular}
\end{table}

\begin{figure}[h]
\centering
\includegraphics[width=0.4\textwidth]{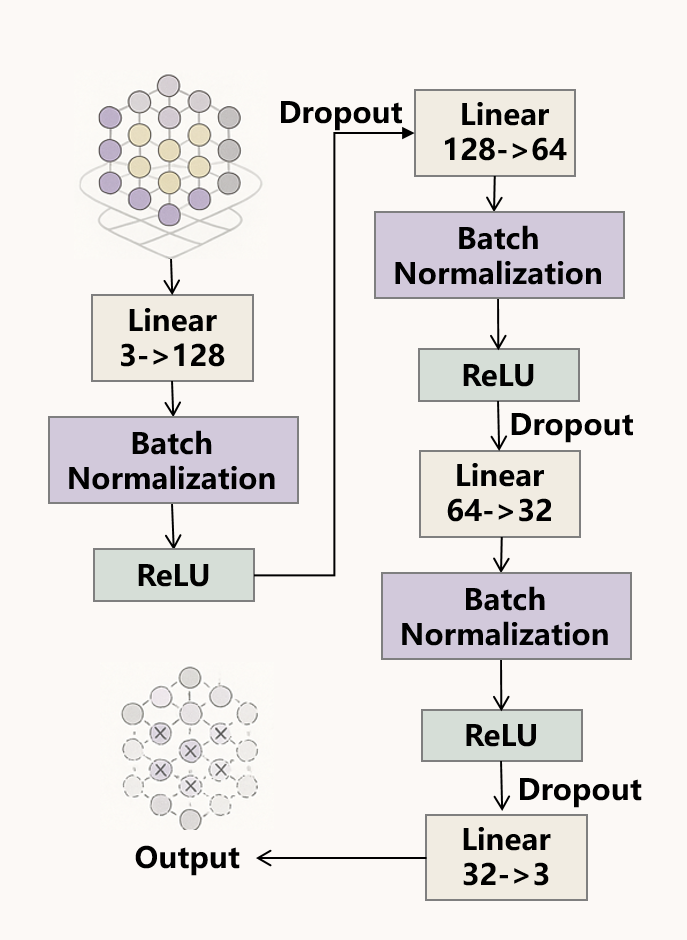}
\caption{Network Structure of the MLP Classifier for Interface-related PIF Modeling}\label{mlp_structure}
\end{figure}

To model the relationship between the proposed metric and PIF targets, we employed a multi-layer perceptron (MLP) classifier composed of four fully connected layers. The input layer accepts a three-dimensional feature vector and maps it to a 128-dimensional hidden representation. This is followed by two intermediate hidden layers with 64 and 32 neurons, respectively, and a final output layer with three neurons corresponding to the classification categories. Each hidden layer is followed by batch normalization to improve training stability and accelerate convergence, a ReLU activation function to introduce non-linearity, and a dropout layer with a probability of 0.3 to prevent overfitting. The structure of the proposed multi-layer perceptron (MLP) classifier is illustrated in Figure~\ref{mlp_structure}.

Given the limited size of the available dataset, we adopted a k-fold cross-validation approach to ensure robust model evaluation and mitigate overfitting. Specifically, a 5-fold cross-validation was performed, yielding classification accuracies of 1.0000, 0.6250, 0.7500, 1.0000, and 1.0000 across the respective folds. The resulting average accuracy was 0.8750, with a standard deviation of ±0.1581. The highest accuracy achieved among all folds was 1.0000, indicating the model's potential to generalize well under certain data partitions.

\begin{figure}[h]
\centering
\includegraphics[width=0.9\textwidth]{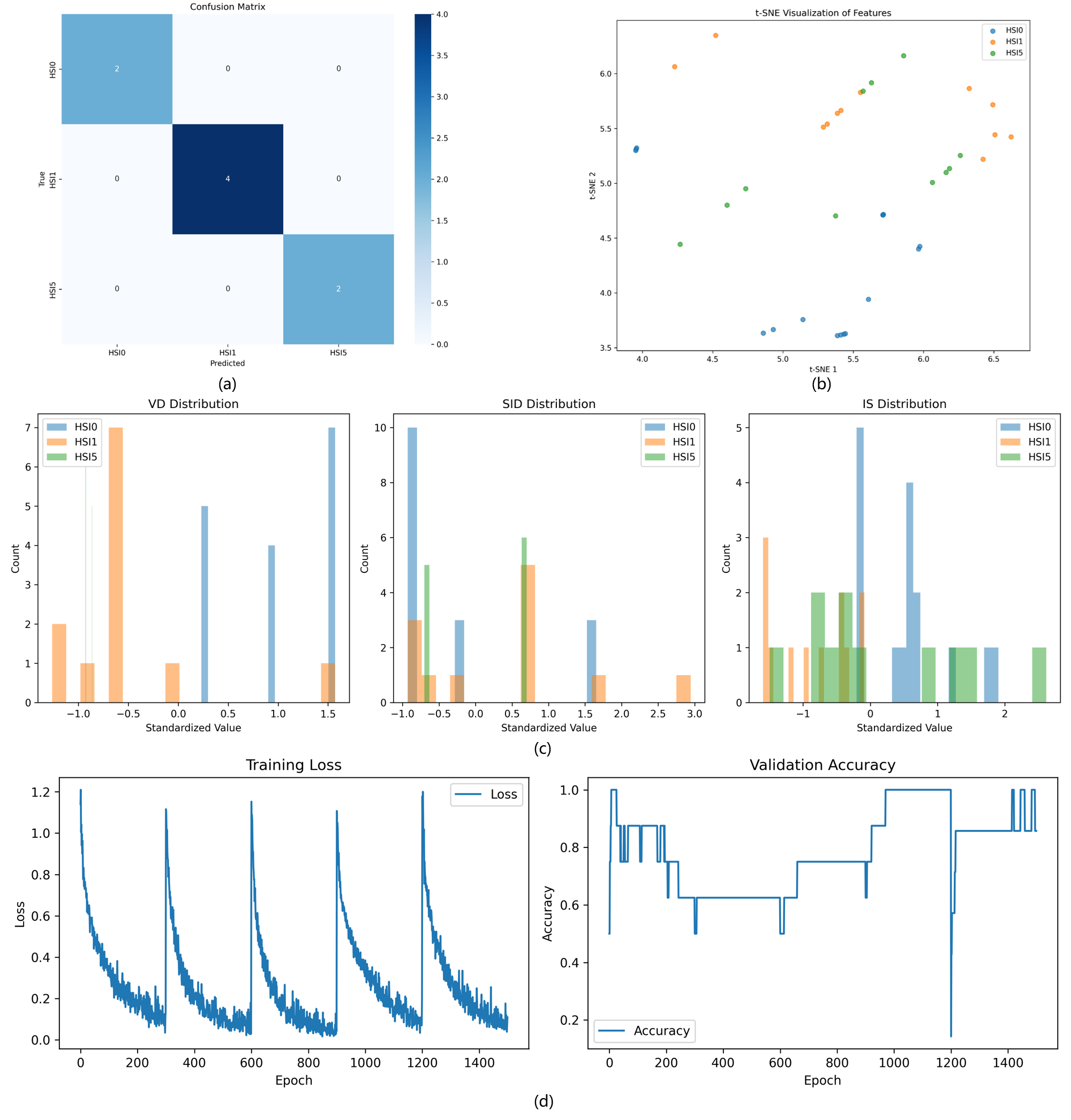}
\caption{Comprehensive Performance Evaluation of the MLP Classifier Including Confusion Matrix, Feature Visualization, and Training Dynamics}\label{model_evaluation}
\end{figure}

Figure~\ref{model_evaluation} provides an integrated assessment of the classifier’s performance across four complementary perspectives. The confusion matrix in Subfigure (a) indicates that the model achieved perfect classification for all instances in the HSI0, HSI1, and HSI5 categories, with 2, 4, and 2 samples, respectively, all correctly predicted. This result highlights the model’s strong predictive capability and the clear separability among the target classes. Subfigure (b) presents the two-dimensional projection of the high-dimensional feature space using the t-distributed Stochastic Neighbor Embedding (t-SNE) technique. The visualization reveals distinct and compact clusters for each class, with limited overlap, particularly between HSI0 and HSI5, thus reinforcing the conclusions drawn from the confusion matrix and suggesting that the model effectively captures discriminative patterns. In Subfigure (c), the standardized distributions of three key input features, visual density (VD), semantic interference density (SID), and interaction spread (IS), are shown across the three classes. VD and IS demonstrate clear class-dependent separation, particularly between HSI0 and HSI1, whereas SID exhibits some overlap yet still contributes to partial discrimination. These patterns confirm the relevance of the selected features in enhancing model performance, especially for HSI0 classification. Finally, Subfigure (d) displays the training dynamics, with training loss decreasing and validation accuracy increasing steadily across epochs, ultimately reaching a high and stable level. These trends align with the previous observations and further support the model’s strong generalization capability, potentially facilitated by favorable class structure and data separability within the given dataset.

\subsection{Designer–User Conflict Analysis and Interface Optimization}

Human–machine conflicts are known to increase operator workload, which in turn elevates the likelihood of human error. Fundamentally, such conflicts often originate from discrepancies between the mental models of system designers and end users—essentially, conflicts between individuals rather than merely between human and machine. However, the relationship between designer-user conflict and human error remains conceptually critical yet frequently misunderstood. 

Based on our experimental findings, we identify three key insights that clarify this relationship. First, the presence of designer-user conflict does not necessarily lead to human error. For instance, in the case of Path $I_{325}$, the calculated PIF was classified as HSI5, indicating a poorly designed interface with significant misalignment between the designer’s intent and user expectations. Nonetheless, participants with strong cognitive adaptability successfully understood the interface function and completed the task without error. This illustrates that conflict may increase cognitive load and reduce operational efficiency but does not invariably result in error. Second, not all observed errors are attributable to designer-user conflict. For example, Path $P_{122}$ was associated with a low-complexity PIF value of HSI1, suggesting an adequately designed interface. However, errors occurred due to operator-specific factors such as fatigue, stress, or inattention-indicating that human error can arise independently of design flaws. Third, although not deterministic, designer–user conflicts do statistically increase the probability of error. In our study, two of the three outcome errors occurred in tasks associated with high-complexity, conflict-prone interfaces (PIF = HSI5), specifically on Paths $P_{216}$ and $P_{323}$. This 66\% occurrence rate highlights a strong correlation between elevated design complexity, inferred conflict, and the likelihood of erroneous outcomes.

\begin{figure}[h]
\centering
\includegraphics[width=0.8\textwidth]{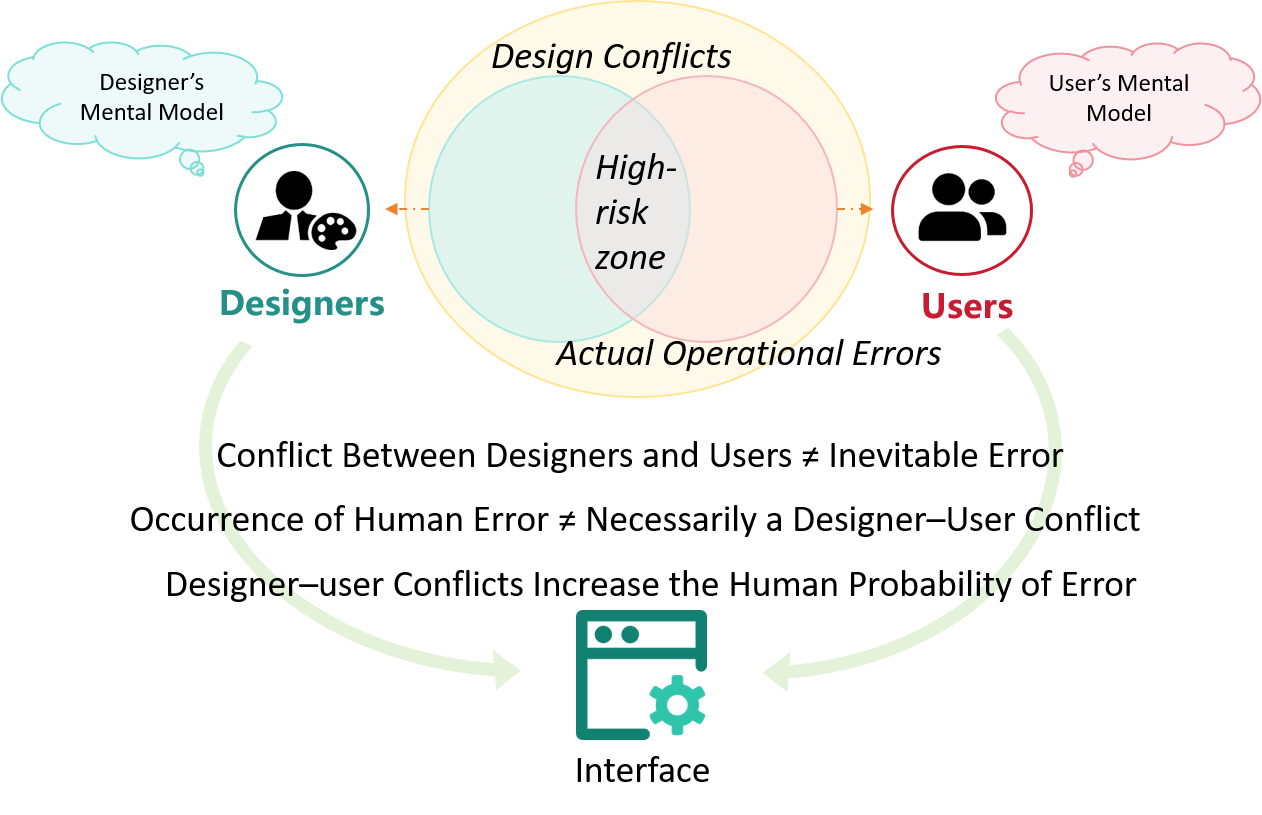}
\caption{Relationship Between Designer-User Conflict and Human Error in Human-System Interaction}\label{Designer}
\end{figure}

The conceptual framework summarizing our findings on designer-user conflict is illustrated in Figure~\ref{Designer}. Both designers and users possess distinct mental models; when these models diverge, design conflicts emerge. The left circle represents the presence of interface or interaction design conflicts-such as ambiguous labeling, non-intuitive navigation structures, or inconsistent affordances, which indicate a misalignment between the designer’s intentions and the user’s expectations. The right circle denotes actual operational errors, including incorrect actions, omissions, or deviations from the prescribed task sequence. The intersection of these two circles captures instances where design conflicts directly contribute to operational errors, representing the highest-risk zone in which both flawed interface design and user behavior jointly undermine system reliability. The left-only region reflects scenarios where design conflicts are present, yet users are able to compensate through prior experience or adaptive strategies, thus maintaining correct performance despite increased cognitive demand. Conversely, the right-only region corresponds to errors that occur independently of design flaws, often driven by human-related factors such as fatigue, distraction, or situational stress, highlighting variability in human performance. Finally, the area outside both circles represents the ideal condition, where neither design conflicts nor operational errors are observed, indicating optimal interface usability and alignment between task demands and human capabilities.

\subsection{Risk-Informed Evaluation Results Based on Predicted PIF Levels}
Furthermore, based on the proposed Insight-R framework, quantitative interface evaluation metrics can be used to generate risk-informed recommendations for the design of new interfaces, even in the absence of experimental validation. As illustrated in Table~\ref{test}, the framework was applied to a set of previously untested procedures, denoted as $TP_{1}$ through $TP_{3}$, to assess their potential usability risks. Specifically, $TP_{1}$ involves verifying whether the outlet feedwater temperature of High-Pressure Heater 1 B in the conventional island system is 190.5 °C; $TP_{2}$ requires checking whether the rotational speed of electric feedwater pump 11 is 5185.7 rpm; and $TP_{3}$ focuses on determining whether the inlet pressure of the same pump is 1.11 MPa. These cases demonstrate the framework’s capability to proactively identify design-induced risks and guide improvements in interface layout and information architecture before real-world deployment.

\begin{table}[h]
\centering
\renewcommand{\arraystretch}{1.3} 
\begin{tabular}{>{\raggedright\arraybackslash}p{1.5cm} >{\raggedright\arraybackslash}p{1.5cm}>{\raggedright\arraybackslash}p{1.5cm}>{\raggedright\arraybackslash}p{2.5cm}>{\raggedright\arraybackslash}p{2.5cm}} 
\hline
\textbf{ID} & \textbf{VD}  & \textbf{SID} & \textbf{IS} & \textbf{Predicted PIF}
\\ 
\hline
$TP_{1}$ & 1/33 & 1/32 & 773.94/2654.05 & HSI1\\
$TP_{2}$ & 1/33 & 4/32 & 823.29/2654.05 & HSI5\\
$TP_{3}$ & 1/33 & 5/32 & 1044.23/2654.05 & HSI5\\
\hline
\end{tabular}
\caption{Predicted Performance Influencing Factor Levels for Untested Conventional Island Operation Procedures}
\label{test}
\end{table}

Using the deep learning-based approach proposed in Section~\ref{Integration with Traditional HRA Methods}, the performance influencing factor (PIF) levels for the untested procedures were computed, as shown in Table~\ref{test}. For $TP_{1}$, the resulting PIF was classified as HSI1, indicating a relatively low risk of human error. In contrast, both $TP_{2}$ and $TP_{3}$ were assigned a PIF level of HSI5, suggesting a higher likelihood of increased Human Error Probability (HEP). These results imply potential usability concerns that may warrant interface redesign. Based on the underlying metrics, risk-informed recommendations include: reducing the number of interactive elements on the interface to lower cognitive load; enhancing semantic differentiation between critical parameters (e.g., the rotational speed and inlet pressure of Electric Feedwater Pump 11); and improving the spatial arrangement to support more intuitive access and recognition. These modifications aim to mitigate operator confusion and reduce the probability of human error under operational conditions.

\section{Conclusions} \label{Conclusion and Discussion}

This study proposes InSight-R, a novel human reliability analysis (HRA) framework driven by an interface-element knowledge graph (IE-KG), aimed at enhancing the objectivity, reproducibility, and interpretability of human failure identification in digital nuclear control rooms. The main contributions and findings are summarized as follows:

\begin{itemize}
    \item A structured, graph-based framework was introduced to enable IE-KG-driven identification of human failure events (HFEs) by mapping empirical data, including operational sequences, errors, and timing, onto interface elements.
    \item A quantitative modeling approach for performance influencing factors (PIFs) was developed and validated, focusing particularly on those related to interface complexity and design constraints.
    \item The proposed framework supports risk-informed interface evaluation, offering an evidence-based complement to expert judgment in the context of HRA.
\end{itemize}

In conclusion, the proposed Insight-R framework represents a significant advancement in the field of human reliability analysis (HRA), facilitating a paradigm shift from traditional expert-dominated approaches to mechanisms grounded in cognitive modeling and data-driven inference. By formalizing the relationship between designer–user conflict and the emergence of human error, our work offers both a theoretical foundation and practical methodology for risk-informed evaluation of digital interfaces. This not only enhances the objectivity and scalability of HRA practices but also provides critical insights for the design of future digitalized human–system interfaces, particularly in safety-critical systems. Furthermore, Insight-R holds promise for supporting the evolution from third-generation to next-generation HRA frameworks, capable of adapting to dynamic operational contexts and heterogeneous data sources. Beyond the nuclear domain, the underlying principles and techniques of Insight-R are applicable to a wide range of domains, including aerospace, healthcare, and intelligent transportation systems, where human–automation interaction remains a central concern.

Looking ahead, future research will focus on integrating cognitive architectures such as ACT-R into the framework, with the goal of enhancing its mechanistic interpretability and predictive power. This integration is expected to further advance dynamic and context-sensitive HRA methodologies, particularly in real-time applications and intelligent control environments.

\section*{Declarations}

\subsection{Funding}
The research was supported by a grant from the National Natural Science Foundation of China (Grant No. T2192933) and the Foundation of National Key Laboratory of Human Factors Engineering (Grant No. HFNKL2024W07).
\subsection{Conflict of interest}
The authors declare that they have no known competing financial interests.

\subsection{Author contribution} 
Xingyu Xiao: Conceptualization, Methodology, Software, Formal analysis, Data Curation, Visualization, Validation, Writing- Original draft preparation. Peng Chen: Software, Methodology. Jiejuan Tong: Conceptualization, Formal analysis, Supervision, Writing - Review and Editing. Shunshun Liu: Methodology. Hongru Zhao: Supervision, Writing - Review and Editing. Jun Zhao: Supervision, Writing - Review and Editing. Qianqian Jia: Supervision, Writing - Review and Editing. Jingang Liang: Resources, Supervision, Writing - Review and Editing, Project administration, Funding acquisition. Wang Haitao: Supervision, Writing- Reviewing and Editing.

\noindent

\bigskip

\bibliography{bibilo2}


\begin{thebibliography}{19}
\ifx \bisbn   \undefined \def \bisbn  #1{ISBN #1}\fi
\ifx \binits  \undefined \def \binits#1{#1}\fi
\ifx \bauthor  \undefined \def \bauthor#1{#1}\fi
\ifx \batitle  \undefined \def \batitle#1{#1}\fi
\ifx \bjtitle  \undefined \def \bjtitle#1{#1}\fi
\ifx \bvolume  \undefined \def \bvolume#1{\textbf{#1}}\fi
\ifx \byear  \undefined \def \byear#1{#1}\fi
\ifx \bissue  \undefined \def \bissue#1{#1}\fi
\ifx \bfpage  \undefined \def \bfpage#1{#1}\fi
\ifx \blpage  \undefined \def \blpage #1{#1}\fi
\ifx \burl  \undefined \def \burl#1{\textsf{#1}}\fi
\ifx \doiurl  \undefined \def \doiurl#1{\url{https://doi.org/#1}}\fi
\ifx \betal  \undefined \def \betal{\textit{et al.}}\fi
\ifx \binstitute  \undefined \def \binstitute#1{#1}\fi
\ifx \binstitutionaled  \undefined \def \binstitutionaled#1{#1}\fi
\ifx \bctitle  \undefined \def \bctitle#1{#1}\fi
\ifx \beditor  \undefined \def \beditor#1{#1}\fi
\ifx \bpublisher  \undefined \def \bpublisher#1{#1}\fi
\ifx \bbtitle  \undefined \def \bbtitle#1{#1}\fi
\ifx \bedition  \undefined \def \bedition#1{#1}\fi
\ifx \bseriesno  \undefined \def \bseriesno#1{#1}\fi
\ifx \blocation  \undefined \def \blocation#1{#1}\fi
\ifx \bsertitle  \undefined \def \bsertitle#1{#1}\fi
\ifx \bsnm \undefined \def \bsnm#1{#1}\fi
\ifx \bsuffix \undefined \def \bsuffix#1{#1}\fi
\ifx \bparticle \undefined \def \bparticle#1{#1}\fi
\ifx \barticle \undefined \def \barticle#1{#1}\fi
\bibcommenthead
\ifx \bconfdate \undefined \def \bconfdate #1{#1}\fi
\ifx \botherref \undefined \def \botherref #1{#1}\fi
\ifx \url \undefined \def \url#1{\textsf{#1}}\fi
\ifx \bchapter \undefined \def \bchapter#1{#1}\fi
\ifx \bbook \undefined \def \bbook#1{#1}\fi
\ifx \bcomment \undefined \def \bcomment#1{#1}\fi
\ifx \oauthor \undefined \def \oauthor#1{#1}\fi
\ifx \citeauthoryear \undefined \def \citeauthoryear#1{#1}\fi
\ifx \endbibitem  \undefined \def \endbibitem {}\fi
\ifx \bconflocation  \undefined \def \bconflocation#1{#1}\fi
\ifx \arxivurl  \undefined \def \arxivurl#1{\textsf{#1}}\fi
\csname PreBibitemsHook\endcsname

\bibitem[\protect\citeauthoryear{Reason}{1990}]{reason1990human}
\begin{botherref}
\oauthor{\bsnm{Reason}, \binits{J.}}:
Human error
(1990)
\end{botherref}
\endbibitem

\bibitem[\protect\citeauthoryear{Hollnagel}{1998}]{hollnagel1998cognitive}
\begin{botherref}
\oauthor{\bsnm{Hollnagel}, \binits{E.}}:
Cognitive reliability and error analysis method (cream)
(1998)
\end{botherref}
\endbibitem

\bibitem[\protect\citeauthoryear{Xiao et~al.}{2024}]{xiao2024krail}
\begin{botherref}
\oauthor{\bsnm{Xiao}, \binits{X.}},
\oauthor{\bsnm{Chen}, \binits{P.}},
\oauthor{\bsnm{Qi}, \binits{B.}},
\oauthor{\bsnm{Zhao}, \binits{H.}},
\oauthor{\bsnm{Liang}, \binits{J.}},
\oauthor{\bsnm{Tong}, \binits{J.}},
\oauthor{\bsnm{Wang}, \binits{H.}}:
Krail: A knowledge-driven framework for base human reliability analysis integrating idheas and large language models.
arXiv preprint arXiv:2412.18627
(2024)
\end{botherref}
\endbibitem

\bibitem[\protect\citeauthoryear{Xiao et~al.}{2025}]{xiao2025dynamic}
\begin{botherref}
\oauthor{\bsnm{Xiao}, \binits{X.}},
\oauthor{\bsnm{Qi}, \binits{B.}},
\oauthor{\bsnm{Liu}, \binits{S.}},
\oauthor{\bsnm{Chen}, \binits{P.}},
\oauthor{\bsnm{Liang}, \binits{J.}},
\oauthor{\bsnm{Tong}, \binits{J.}},
\oauthor{\bsnm{Wang}, \binits{H.}}:
A dynamic risk-informed framework for emergency human error prevention in high-risk industries: A nuclear power plant case study.
Reliability Engineering \& System Safety,
111080
(2025)
\end{botherref}
\endbibitem

\bibitem[\protect\citeauthoryear{Xiao et~al.}{2024}]{xiao2024emergency}
\begin{barticle}
\bauthor{\bsnm{Xiao}, \binits{X.}},
\bauthor{\bsnm{Liang}, \binits{J.}},
\bauthor{\bsnm{Tong}, \binits{J.}},
\bauthor{\bsnm{Wang}, \binits{H.}}:
\batitle{Emergency decision support techniques for nuclear power plants: Current state, challenges, and future trends}.
\bjtitle{Energies}
\bvolume{17}(\bissue{10}),
\bfpage{2439}
(\byear{2024})
\end{barticle}
\endbibitem

\bibitem[\protect\citeauthoryear{Xiao et~al.}{2025}]{xiao2025cognitive}
\begin{botherref}
\oauthor{\bsnm{Xiao}, \binits{X.}},
\oauthor{\bsnm{Chen}, \binits{P.}},
\oauthor{\bsnm{Tong}, \binits{J.}},
\oauthor{\bsnm{Liu}, \binits{S.}},
\oauthor{\bsnm{Zhao}, \binits{H.}},
\oauthor{\bsnm{Zhao}, \binits{J.}},
\oauthor{\bsnm{Jia}, \binits{Q.}},
\oauthor{\bsnm{Liang}, \binits{J.}},
\oauthor{\bsnm{Wang}, \binits{H.}}:
A cognitive-mechanistic human reliability analysis framework: A nuclear power plant case study.
arXiv preprint arXiv:2504.18604
(2025)
\end{botherref}
\endbibitem

\bibitem[\protect\citeauthoryear{Park et~al.}{2020}]{park2020inter}
\begin{barticle}
\bauthor{\bsnm{Park}, \binits{J.}},
\bauthor{\bsnm{Jung}, \binits{W.}},
\bauthor{\bsnm{Kim}, \binits{J.}}:
\batitle{Inter-relationships between performance shaping factors for human reliability analysis of nuclear power plants}.
\bjtitle{Nuclear Engineering and Technology}
\bvolume{52}(\bissue{1}),
\bfpage{87}--\blpage{100}
(\byear{2020})
\end{barticle}
\endbibitem

\bibitem[\protect\citeauthoryear{Xing et~al.}{2020}]{xing2020integrated}
\begin{botherref}
\oauthor{\bsnm{Xing}, \binits{J.}},
\oauthor{\bsnm{Chang}, \binits{Y.}},
\oauthor{\bsnm{DeJesus}, \binits{J.}}:
Integrated human event analysis system for event and condition assessment (idheas-eca).
US Nuclear Regulatory Commission, Washington, DC
(2020)
\end{botherref}
\endbibitem

\bibitem[\protect\citeauthoryear{Thimbleby et~al.}{2015}]{thimbleby2015unreliable}
\begin{barticle}
\bauthor{\bsnm{Thimbleby}, \binits{H.}},
\bauthor{\bsnm{Oladimeji}, \binits{P.}},
\bauthor{\bsnm{Cairns}, \binits{P.}}:
\batitle{Unreliable numbers: error and harm induced by bad design can be reduced by better design}.
\bjtitle{Journal of The Royal Society Interface}
\bvolume{12}(\bissue{110}),
\bfpage{20150685}
(\byear{2015})
\end{barticle}
\endbibitem

\bibitem[\protect\citeauthoryear{Roll et~al.}{2019}]{roll2019human}
\begin{barticle}
\bauthor{\bsnm{Roll}, \binits{L.C.}},
\bauthor{\bsnm{Siu}, \binits{O.-l.}},
\bauthor{\bsnm{Li}, \binits{S.Y.}},
\bauthor{\bsnm{De~Witte}, \binits{H.}}:
\batitle{Human error: The impact of job insecurity on attention-related cognitive errors and error detection}.
\bjtitle{International journal of environmental research and public health}
\bvolume{16}(\bissue{13}),
\bfpage{2427}
(\byear{2019})
\end{barticle}
\endbibitem

\bibitem[\protect\citeauthoryear{O'Hara and Fleger}{2020}]{o2020human}
\begin{botherref}
\oauthor{\bsnm{O'Hara}, \binits{J.M.}},
\oauthor{\bsnm{Fleger}, \binits{S.}}:
Human-system interface design review guidelines.
Technical report,
Brookhaven National Lab.(BNL), Upton, NY (United States)
(2020)
\end{botherref}
\endbibitem

\bibitem[\protect\citeauthoryear{Stone et~al.}{2005}]{stone2005user}
\begin{botherref}
\oauthor{\bsnm{Stone}, \binits{D.}},
\oauthor{\bsnm{Jarrett}, \binits{C.}},
\oauthor{\bsnm{Woodroffe}, \binits{M.}},
\oauthor{\bsnm{Minocha}, \binits{S.}}:
User interface design and evaluation
(2005)
\end{botherref}
\endbibitem

\bibitem[\protect\citeauthoryear{Thimbleby}{1990}]{thimbleby1990user}
\begin{botherref}
\oauthor{\bsnm{Thimbleby}, \binits{H.}}:
User interface design
(1990)
\end{botherref}
\endbibitem

\bibitem[\protect\citeauthoryear{Morland}{1983}]{morland1983human}
\begin{barticle}
\bauthor{\bsnm{Morland}, \binits{D.V.}}:
\batitle{Human factors guidelines for terminal interface design}.
\bjtitle{Communications of the ACM}
\bvolume{26}(\bissue{7}),
\bfpage{484}--\blpage{494}
(\byear{1983})
\end{barticle}
\endbibitem

\bibitem[\protect\citeauthoryear{Carlgren et~al.}{2014}]{carlgren2014design}
\begin{barticle}
\bauthor{\bsnm{Carlgren}, \binits{L.}},
\bauthor{\bsnm{Elmquist}, \binits{M.}},
\bauthor{\bsnm{Rauth}, \binits{I.}}:
\batitle{Design thinking: Exploring values and effects from an innovation capability perspective}.
\bjtitle{The Design Journal}
\bvolume{17}(\bissue{3}),
\bfpage{403}--\blpage{423}
(\byear{2014})
\end{barticle}
\endbibitem

\bibitem[\protect\citeauthoryear{Chen and Tong}{2024}]{chen2024influence}
\begin{bchapter}
\bauthor{\bsnm{Chen}, \binits{P.}},
\bauthor{\bsnm{Tong}, \binits{J.}}:
\bctitle{Influence of nuclear power plant interface complexity on operator performance: A modeling study}.
In: \bbtitle{International Conference on Nuclear Engineering},
vol. \bseriesno{88230},
pp. \bfpage{003}--\blpage{03011}
(\byear{2024}).
\bcomment{American Society of Mechanical Engineers}
\end{bchapter}
\endbibitem

\bibitem[\protect\citeauthoryear{Chang et~al.}{2021}]{chang2021idheas}
\begin{bchapter}
\bauthor{\bsnm{Chang}, \binits{Y.J.}},
\bauthor{\bsnm{Xing}, \binits{J.}},
\bauthor{\bsnm{DeJesus~Segarra}, \binits{J.}}:
\bctitle{Idheas suite for human reliability analysis}.
In: \bbtitle{Proceedings of the 2021 International Topical Meeting on Probabilistic Safety Assessment and Analysis (PSA 2021)},
pp. \bfpage{1231}--\blpage{1238}
(\byear{2021})
\end{bchapter}
\endbibitem

\bibitem[\protect\citeauthoryear{Nussbaum et~al.}{2024}]{nussbaum2024nomic}
\begin{botherref}
\oauthor{\bsnm{Nussbaum}, \binits{Z.}},
\oauthor{\bsnm{Morris}, \binits{J.X.}},
\oauthor{\bsnm{Duderstadt}, \binits{B.}},
\oauthor{\bsnm{Mulyar}, \binits{A.}}:
Nomic embed: Training a reproducible long context text embedder.
arXiv preprint arXiv:2402.01613
(2024)
\end{botherref}
\endbibitem

\bibitem[\protect\citeauthoryear{Chang et~al.}{2016}]{chang2016human}
\begin{bchapter}
\bauthor{\bsnm{Chang}, \binits{Y.J.}},
\bauthor{\bsnm{Xing}, \binits{J.}},
\bauthor{\bsnm{Peters}, \binits{S.}}:
\bctitle{Human reliability analysis method development in the us nuclear regulatory commission}.
In: \bbtitle{13th International Conference on Probabilistic Safety Assessment and Management (PSAM 13), Seoul, Korea}
(\byear{2016})
\end{bchapter}
\endbibitem

\end{thebibliography}

\end{document}